\documentstyle[preprint,aps,epsfig]{revtex}
\newcommand{\lsim}{\mathrel{\mathop{\kern 0pt \rlap
  {\raise.2ex\hbox{$<$}}}
  \lower.9ex\hbox{\kern-.190em $\sim$}}}
\newcommand{\gsim}{\mathrel{\mathop{\kern 0pt \rlap
  {\raise.2ex\hbox{$>$}}}
  \lower.9ex\hbox{\kern-.190em $\sim$}}}
\newcommand{\be}     {\begin{equation}}
\newcommand{\ee}     {\end{equation}}
\newcommand{\bea}     {\begin{eqnarray}}
\newcommand{\eea}     {\end{eqnarray}}
\begin{document}
\preprint{
\vbox{ \hbox{SNUTP\hspace*{.2em}01-010}
}}
\title{ 
Muon anomalous  magnetic
moment 
and the stabilized Randall-Sundrum scenario  
}

\author{
Seong Chan Park\footnote{spark@fire.snu.ac.kr} 
~and ~~ H.~S.~ Song\footnote{hssong@physs.snu.ac.kr} 
}
\vspace{1.5cm}
\address{
School of Physics, Seoul National University,
Seoul 151-742, Korea
}

\maketitle
\thispagestyle{empty}
\setcounter{page}{1}

\begin{abstract}
\noindent 
We study the effects of extra dimension on the muon anomalous
magnetic moment  
in the stabilized Randall-Sundrum scenario. 
The effects of the Kaluza-Klein states heavier than the 
cut-off scale expected to be of order $\Lambda_\pi$ are
neglected.   
Contribution from the spin-2 Kaluza-Klein states dominates
over that from the spin-0 radion.
The recent BNL E821 results impose a strict constraint
on the parameter space of the model: 
$\Lambda_\pi \approx 0.4 - 2.1$ TeV
with $k/ M_{\rm Pl} =0.01 - 0.1$.  
Small $k/M_{\rm Pl}$ is
preferred if $\Lambda_\pi$ is TeV scale.

\end{abstract}
\vskip 0.5cm
\noindent
PACS number(s): 11.30.Pb, 11.30.Er

\newpage

Recent BNL E821 measurement of the anomalous magnetic moment
of the muon {\cite {BNL}} has drawn big interest 
since the result shows $2.6 \sigma$ deviation
from the standard model (SM) prediction and may be indeed
a harbinger of the new physics at TeV scale{\cite {Marciano}}
\be
\delta a_\mu \equiv a_\mu ({\rm exp}) - a_\mu ({\rm SM}) = 43(16) \times 10^{-10}
\ee
where $a_\mu \equiv (g_\mu -2)/2$ and $a_\mu ({\rm exp})$ is the world
average of experimental values.
Since much more data have been already obtained and statistical 
improvements of the experiment will be achieved, it is timely to see
the implication of the advocates of the new physics on $a_\mu$.
Many supersymmetric{\cite {susy}} and non-supersymmetric{\cite {nonsusy}}
 interpretations for this anomaly were already proposed,
and the combined constraints with other experiments and observations
were studied in the particular models{\cite {ellis}}.

In this study, 
we consider
the implication of extra dimensional model suggested by 
Randall and Sundrum (RS)
{\cite {Randall:1999ee}}.
RS presented a 5-dimensional non-factorizable geometry, based on 
$AdS_5$ spacetime, to generate the hierarchy. In this case, the weak
scale could be geometrically generated from the warp factor which arises
from the well adjusted boundary conditions reconciled with four dimensional
Poincar${\rm \acute{e}}$ invariance on the branes. 
The extra dimension is an interval with 
a $S^1/Z_2$ orbifold symmetry 
and two 3 branes with opposite tensions are located
at fixed points at $\phi=0$ and $\phi=\pi$ where
$\phi$ is the angular coordinate of the fifth dimension.  
From the four dimensional effective action in RS scenario, one can get
the relation between the induced four dimensional Planck scale $M_{\rm Pl}$
and the fundamental quantum gravitation/string scale $M_S$;
\be
M_{\rm Pl}^2=\frac{M_S^3}{k}(1-e^{-2kr_c \pi})    
\ee
where $k$ is $ AdS_5$
curvature and $r_c$ is the radius of extra orbifold dimension. 
Assuming that we live on the brane with a negative tension, a field on
this brane with the fundamental mass $m_0$ will appear to have the physical
mass $m=e^{-kr_c\pi}m_0$. Thus the weak scales are generated via the
exponential warp factor from the Planck scale with $kr_c \simeq 12$. 
The masses of the
KK graviton and their couplings to the SM fields could be determined by
mode expansions and proper normalizations. 
For the n-th excited state the mass is given as 
\cite{Goldberger:1999wh},
\be
m_n = k x_n e^{-kr_c \pi},
\ee
where $x_n$ is a root of the Bessel function of order 1, 
$k/M_{\rm Pl}$ is $AdS_5$ curvature in the unit of Planck scale
and $\Lambda_\pi = e^{-kr_c\pi}M_{\rm Pl}\sim {\rm TeV}$ is effective weak 
scale. It should be noted that, we impose the condition $k<M_{\rm pl}$ 
so that we trust RS solution \cite{Randall:1999ee}.
Thus, $k/M_{\rm Pl}$ must be smaller than 1 and we take the value in the 
range of $0.01 \leq \ k/M_{\rm Pl} \leq 1$ as conservative bound. Then, the lowest
KK graviton has mass slightly larger than $1$ TeV for 
$\Lambda_\pi \sim 3$ TeV and so 
there might be a chance to see the effects of KK graviton 
at near future colliders. 
  
The interaction Lagrangians for the zeroth KK graviton mode and standard
model fields are suppressed by the Planck scale. But
the excited modes interact with SM field through 
the couplings suppressed by the TeV scale,
$\Lambda_\pi$;   

\be
{\cal L}= -\frac{1}{M_{\rm Pl}} T_{\mu\nu}({\rm SM})h_0^{\mu\nu}
          -\frac{1}{\Lambda_\pi}\sum_n T_{\mu\nu}({\rm SM})h_n^{\mu\nu}.  
\ee
Here note that the mass of the KK states could become 
as large as Planck scale or infinity if there were no physical cut-off
scale. These super massive KK states could affect the calculations of physical 
processes or quantities. 
However, since we do not perfectly know the ultimate
theory of quantum gravity or the relevant string theory, their effects
could not be understood by a way of fully controllable methods yet. 
Henceforth the interaction Lagrangian 
given above will be understood as an effective
Lagrangian with the cut-off $\Lambda$ 
expected to be of order $\Lambda_\pi$ \cite{Davoudiasl} so that
the effects of the KK states above the cut-off scale 
are assumed to be integrated out in our analysis.      

On the other hand, one can get the large value of $kr_c \simeq 12$ through
the moduli stabilization mechanism proposed by Goldberger and Wise(GW)
\cite{Goldberger:1999uk} without fine-tuning when the small bulk scalar mass
is assumed. As a consequence, mass of the radion($m_\phi$) would be a 
order smaller than TeV scale because it is  suppressed not only
by warp factor but also by the small ratio of the bulk mass of the radion 
and the curvature $k$. 
For typical arrangement in GW mechanism, the mass of radion would be 
$m_\phi \simeq 0.06 \Lambda_\pi \leq 180~{\rm GeV}$ for $\Lambda_\pi \leq 3~{\rm TeV}$ 
\cite{Csaki:2000zn}.
Note that the mass of radion would be smaller than that of excited KK graviton.
Thus the radion could provide first chance to observe RS-scenario.
The radion couples to ordinary matter through the trace of the SM energy
momentum tensor $T_\mu^\mu({\rm SM})$ with TeV scale suppressed coupling ;
\be
{\cal L}=\frac{\phi}{\Lambda_\phi}T_\mu^\mu({\rm SM}),
\ee
where $\Lambda_\phi=\sqrt{6}\Lambda_\pi$ is the vacuum expectation value for 
the radion field\cite{Giudice:2000av}.

The anomalous magnetic moment of the muon is the coefficient of the operator
$(e/4 m_\mu) \bar{\mu} \sigma^{\alpha \beta} \mu F_{\alpha \beta}$,
where $\sigma^{\alpha \beta}=\frac{i}{2}[\gamma^\alpha, \gamma^\beta]$
is the Lorentz generator for spin-1/2 spinors. This term can be calculated
by considering loop induced correction to $\mu\mu\gamma$ vertex. 
The extra dimensional contribution of the RS model, 
to the anomalous magnetic moment $\delta a^{\rm RS}_\mu$,
is dominated by one-loop diagrams with either the tower of
Kaluza-Klein(KK) spin-2 graviton or radion. 
Since the zeroth KK-mode of graviton is just massless graviton with
the Planck scale suppressed coupling, the dominant contribution would come 
from the KK excited states. 

It is convenient to express the contribution of 
the KK gravitons as (See Fig.1)
\be
\delta a^{KK}= \frac{1}{16\pi^2} (\frac{m_\mu}{\Lambda_\pi})^2 {\cal A}
\ee
where ${\cal A}$ is essentially effective degree of freedom and
it could be obtained by summing over all relevant contributions 
of all the relevant KK states in the Feynman diagrams.
The muon mass is $m_\mu$ and the factor$\frac{1}{16\pi^2}$ 
is the usual loop factor. 
By the gravitational Ward identity, we expect that those terms in the
numerator of the graviton propagator containing two or more external
momenta do not contribute to $a_\mu$ at the order 
$\sim {\cal O}(\frac{m_\mu}{\Lambda_\pi})^2$.
The sum of the partial amplitudes at the limit of 
$(\frac{m_\mu}{m_{KK}})^2 \rightarrow 0$ for our interest,
\be
{\cal A} \approx 5 \times N_\Lambda,
\ee
where $N_\Lambda$ denotes the number of KK states up to cut-off $\Lambda$.
We can understand the factor 5 as the number of 
spin degrees of freedom for each massive spin-2 KK state and it is 
shown by explicit calculation. 
Since the universality of the gravitational coupling is preserved
in this work, the one-loop finiteness of the gravitational 
coupling appears as the case for massive {\cite{Graesser}}
and massless graviton \cite{massless}. 
This result is different from the case of \cite{Davoudiasl}
where the bulk gauge and fermion fields are assumed to contribute
to $a_\mu$.
As was commented in \cite{Graesser}, The KK-states exhibit 
non-decoupling properties 
since their couplings to matter are stronger at higher
energies.
But we get the finite contribution with finite $N_\Lambda$
which is related as 
$(k/M_{\rm Pl})\Lambda_\pi (N_\Lambda \pi +\pi/4 ) \approx \Lambda \approx \Lambda_\pi$.
We use the asymptotic formula for the zeroes of the Bessel function of
order 1: $x_N \approx N\pi+\pi/4$ at large $N$. 
The whole KK contribution is given by
\bea
\delta a^{KK} &\approx& \frac{5}{16\pi^2}
(\frac{\lambda}{\pi}\frac{M_{\rm Pl}}{k}-\frac{1}{4})
(\frac{m_\mu}{\Lambda_\pi})^2  
\eea
with $\lambda \equiv \Lambda /\Lambda_\pi \sim {\cal O}(1)$.

The dominant radion contribution to the muon anomaly arises from the
only one diagram and $\delta a^\phi$ is given by{\cite {mahanta}}
\be
\delta a^\phi 
= \frac{1}{4\pi^2}(\frac{m_\mu}{\Lambda_\phi})^2
 \int dx \frac{x^2 (2-x)}{x^2 + R(1-x)}  
\leq  \frac{1}{16\pi^2}(\frac{m_\mu}{\Lambda_\pi})^2, 
\ee   
where  $R\equiv (\frac{m_\phi}{m_\mu})^2 $ is in general free parameter.
The upper bound is saturated at the limit of $R \rightarrow 0$ and
$\delta a^\phi$ decrease with larger $R$. 
Thus $\delta a^\phi$ is
much smaller than $\delta a^{KK}$ in the preferred 
parameter space of the model.
It is worth noting that the bound from muon
anomaly experiments are relevant only for very light radion. But we
do not consider that possibility here.

Finally, the total contribution to $a_\mu$ from the both of KK-modes and
the radion could be approximated as
\be
\delta a^{\rm RS}_\mu \approx\delta a^{KK} 
\approx 
8.8 \times 10^{-11} 
\times (\frac{k}{M_{\rm Pl}})^{-1}
\times (\frac{\rm TeV}{\Lambda_\pi})^2 
\ee   
where $\lambda$ is set as $\pi/4$ to be order 1.   
If we ascribe the difference $a_\mu ({\rm exp}) -a_\mu ({\rm SM})$
to $\delta a^{\rm RS}$, for $k/ M_{\rm Pl}$ in the range
0.01 - 0.1, then $\Lambda_\pi \approx 0.4 - 2.1$ TeV.
The Fig.2 describes the allowed parameter space $(\Lambda_\pi, k/M_{\rm Pl})$
for the muon anomaly. The solid line and the dotted line denotes the
lower and upper bound of the experimental value, respectively. 
We can see that the small $k/ M_{\rm Pl}$ 
is preferred with TeV scaled $\Lambda_\pi$. 
The large range of the preferred parameter space compensating $\delta a_\mu$
data is already disfavored from the
experiments of Tevatron or LEP-II{\cite {colliders}}.
But we still don't dispute the possibility of RS model appearing at
over TeV scale.  
So we are eagerly waiting for 
the expected improved measurements which will give much more strict
constraints on the model.

In summary, we have studied the effects of Kaluza-Klein tower and radion
on the $g-2$ factor of muon in the stabilized RS model at 
one loop level. We discarded the effects of the KK states heavier than
the cut-off scale.
The Kaluza-Klein states are shown to affect much more greatly 
than the radion do on the muon anomalous magnetic moment.  
The recently reported deviation in anomalous 
magnetic moment is possibly accommodated in the 
corresponding parameter
space in the range of $\Lambda_\pi \approx 0.4 - 2.1$ GeV
with $k/ M_{\rm Pl} =0.01 - 0.1$. Small $k/M_{\rm Pl}$ is
preferred if $\Lambda_\pi$ is TeV scale.  

\acknowledgments
We thank K. Y. Lee and
Jung-Tay Yee for useful discussion, J. Song for valuable criticism. 
The work was supported by the BK21 program. 


%
\begin{figure}[htb]
\vskip 2.5cm  
\begin{center}
 \hbox to\textwidth{\hss\epsfig{file=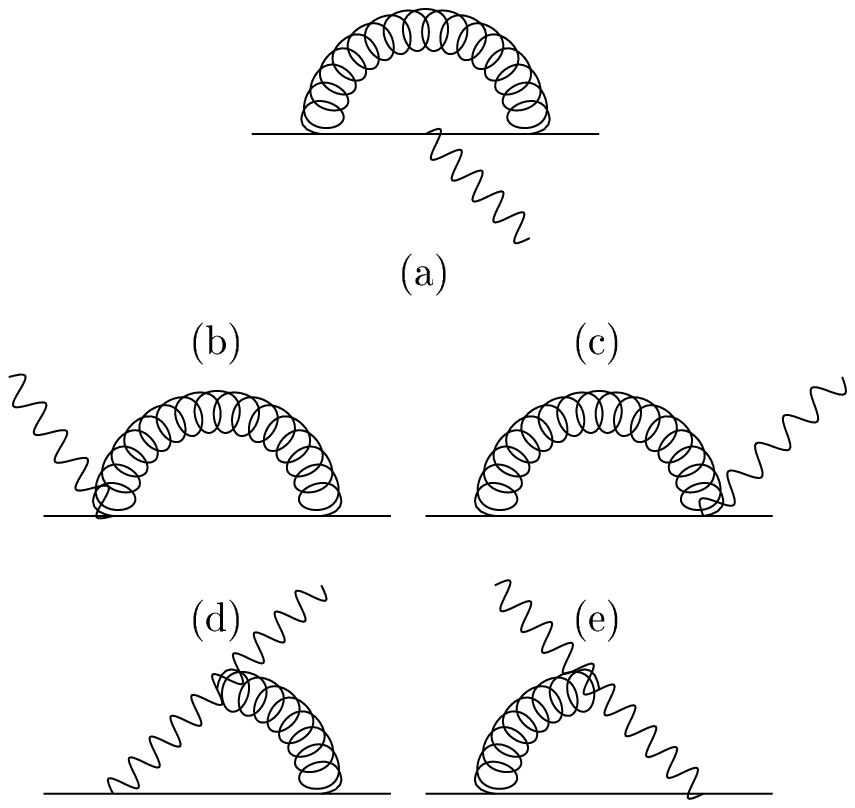,width=4cm,
                    height=4cm}\hss}
 \end{center}
 \vskip 0.5cm
\caption{\it The Kaluza-Klein mediated 1-loop diagrams for (g-2)of the muon 
. The solid and wavy line denotes muon and photon on-shell. 
The spring-shape lines denote spin-2 Kaluza-Klein states. } 
 \label{fig:fig1}
\end{figure}

\begin{figure}[htb]
\begin{center}
 \hbox to\textwidth{\hss\epsfig{file=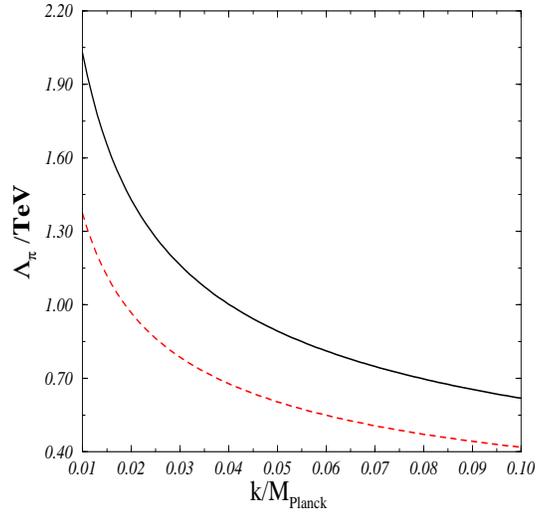,width=7cm,
                    height=7cm}\hss}
 \end{center}
 \vskip 0.5cm
\caption{\it 
The allowed parameter space ($\Lambda_\pi ,k/M_{\rm Pl}$) to
compensating recent BNL E821 result of the muon anomalous
magnetic moment. The (black)solid line and the (red) dotted line
denotes lower and upper limit of the muon anomaly ,respectively.  
} 
 \label{fig:fig2}
\end{figure}

\end{document}